\documentclass[a4paper,12pt]{article}

\usepackage{CJK}
\usepackage{mathtools}
\usepackage{tikz}
\usepackage{amssymb}
\usepackage{epsfig}
\usepackage{bbm}
\usepackage{bbold}
\usepackage{tabularx,booktabs}
\usepackage{graphicx}
\usepackage{hyperref}
\usepackage{graphics,graphpap}
\usepackage{graphicx}
\usepackage{amsmath,exscale,amsthm,bm}
\usepackage{graphicx}
\usepackage{yfonts}
\usepackage{mathrsfs}
\usepackage[toc]{appendix}
\usetikzlibrary{trees}
\usetikzlibrary{positioning,arrows}
\usetikzlibrary{decorations.pathmorphing}
\usetikzlibrary{decorations.markings}
\usetikzlibrary{decorations.text}
\usetikzlibrary{fit,chains,calc,shapes.geometric,intersections}
\usepackage{natbib}
\setcitestyle{square,comma,numbers,sort&compress}
\usepackage{circledsteps}
\def\mc#1{\mathcal#1}

\allowdisplaybreaks[1]

\makeatletter

\newcommand{\Rmnum}[1]{\expandafter\@slowromancap\romannumeral #1@}
\makeatother

\begin{document}

\title{
\vspace*{15mm}
\bf The split majoron model confronts the NANOGrav signal and cosmological tensions} 
\author{{\Large Pasquale~Di~Bari$^{a}$, Moinul Hossain Rahat$^{a,b}$}
\\
$^a$
{\it\small School of Physics and Astronomy},
{\it\small University of Southampton,}
{\it\small  Southampton, SO17 1BJ, U.K.}\\
$^b$
{\it\small Instituto de F\'isica Corpuscular, Universidad de Valencia},  \\
{\it\small and CSIC, Edificio Institutos Investigaci\'on, C/Catedr\'atico},
{\it\small  Jos\'e Beltr\'an 2, 46980 Paterna, Spain}
}

\maketitle \thispagestyle{empty}

\abstract{
In the light of the evidence of a gravitational wave background from the NANOGrav 15yr data set, we
reconsider the split majoron model as a new physics extension of the standard model able to
generate a needed contribution to solve the current tension between the data and the standard interpretation
in terms of inspiraling supermassive black hole massive binaries. In the split majoron model 
the  seesaw right-handed neutrinos acquire Majorana masses from spontaneous  symmetry breaking of global $U(1)_{B-L}$ 
in a strong first order phase transition of a complex scalar field occurring above the electroweak scale.  The final 
vacuum expectation value couples to a second complex scalar field undergoing a low scale phase transition occurring
after neutrino decoupling. Such a coupling enhances the strength of this second low scale first order phase transition and can 
generate a sizeable primordial gravitational wave background contributing to the NANOGrav 15yr signal. Some amount of extra-radiation is generated after neutron-to-proton ration freeze-out but 
prior to nucleosynthesis. This can be either  made compatible with current 
upper bound from primordial deuterium measurements or even be used to solve a potential deuterium problem.
Moreover, the free streaming length of light neutrinos can be 
suppressed by their interactions with the resulting majoron background 
and this mildly ameliorates existing cosmological tensions. 
Thus cosmological observations nicely provide independent motivations for the model.  
}

\maketitle
\flushbottom

\def\a{\alpha}
\def\b{\beta}
\def\c{\chi}
\def\d{\delta}
\def\e{\epsilon}
\def\f{\phi}
\def\g{\gamma}
\def\h{\eta}
\def\i{\iota}
\def\j{\psi}
\def\k{\kappa}
\def\la{\lambda}
\def\m{\mu}
\def\n{\nu}
\def\o{\omega}
\def\p{\pi}
\def\q{\theta}
\def\r{\rho}
\def\s{\sigma}
\def\t{\tau}
\def\u{\upsilon}
\def\x{\xi}
\def\z{\zeta}
\def\D{\Delta}
\def\F{\Phi}
\def\G{\Gamma}
\def\J{\Psi}
\def\L{\Lambda}
\def\O{\Omega}
\def\P{\Pi}
\def\Q{\Theta}
\def\S{\Sigma}
\def\U{\Upsilon}
\def\X{\Xi}

\def\ve{\varepsilon}
\def\vf{\varphi}
\def\vr{\varrho}
\def\vs{\varsigma}
\def\vq{\vartheta}

\newcommand{\vev}[1]{\langle #1 \rangle}
\def\dg{\dagger}                                     
\def\ddg{\ddagger}                                   
\def\wt#1{\widetilde{#1}}                    
\def\mt{\widetilde{m}_1}
\def\mti{\widetilde{m}_i}
\def\mtj{\widetilde{m}_j}
\def\rt{\widetilde{r}_1}
\def\mtt{\widetilde{m}_2}
\def\mttt{\widetilde{m}_3}
\def\rtt{\widetilde{r}_2}
\def\mb{\overline{m}}
\def\VEV#1{\left\langle #1\right\rangle}        
\def\be{\begin{equation}}
\def\ee{\end{equation}}
\def\ds{\displaystyle}
\def\ra{\rightarrow}

\def\bea{\begin{eqnarray}}
\def\eea{\end{eqnarray}}
\def\NO{\nonumber}
\def\Bar#1{\overline{#1}}
\def\ylz{\textcolor{red}}

\section{Introduction}

The NANOGrav collaboration has found evidence for a gravitational wave (GW) background at $\sim$ nHZ frequencies 
in the 15-year data set \cite{NANOGrav:2023gor, NANOGrav:2020bcs, NANOGrav:2021flc, NANOGrav:2023hvm}.
This strongly relies on the observed correlations among 67 pulsars following an expected Hellings-Downs pattern for a 
stochastic GW background \cite{Hellings:1983fr}. A simple baseline model is provided by a standard interpretation in terms of
inspiraling supermassive black hole binaries (SMBHBs) with a fiducial $f^{-2/3}$ characteristic strain spectrum.   Such a baseline model
provides a poor fit to the data and some deviation is currently favoured. 
In particular, the collaboration finds that models where in addition
to SMBHBs one also has a contribution from new physics, provide a better fit to the 
NANOGrav data than the baseline model, resulting in Bayes factors between 10 and 100 \cite{NANOGrav:2023hvm}\footnote{see Refs.~\cite{Yi:2023mbm, Kuang:2023ygc, Figueroa:2023zhu, Unal:2023srk, Abe:2023yrw, Cacciapaglia:2023kat, Anchordoqui:2023tln, Li:2023bxy, Xiao:2023dbb, Lu:2023mcz, Zhang:2023lzt, Konoplya:2023fmh, Chowdhury:2023opo, Niu:2023bsr, Liu:2023ymk, Westernacher-Schneider:2023cic, Gouttenoire:2023nzr, Ghosh:2023aum, Datta:2023vbs, Borah:2023sbc, Barman:2023fad, Bi:2023tib, Wang:2023ost, Broadhurst:2023tus, Yang:2023qlf, Eichhorn:2023gat, Huang:2023chx, Gouttenoire:2023ftk, Cai:2023dls, Inomata:2023zup, Lazarides:2023ksx, Depta:2023qst, Blasi:2023sej, Bian:2023dnv, Franciolini:2023wjm, Shen:2023pan, Lambiase:2023pxd, Han:2023olf, Guo:2023hyp, Wang:2023len, Ellis:2023tsl, Vagnozzi:2023lwo, Fujikura:2023lkn, Kitajima:2023cek, Franciolini:2023pbf, Megias:2023kiy, Ellis:2023dgf, Bai:2023cqj, Yang:2023aak, Ghoshal:2023fhh, Deng:2023btv, Athron:2023mer, Addazi:2023jvg, Oikonomou:2023qfz, Kitajima:2023vre, Mitridate:2023oar, King:2023cgv, Liu:2023hte} for some recent new physics approaches.}. 
 
First order phase transitions at low scales could potentially provide such an additional contribution. For temperatures of the phase transition
in the range 1 MeV -- 1 GeV, the resulting GW background may explain the entire NANOGrav signal \cite{NANOGrav:2023gor, NANOGrav:2023hvm}. 
However, when a realistic model is considered, one needs also to take into account the cosmological constraints on the amount of 
extra radiation from big bang nucleosynthesis (BBN) and CMB anisotropies.  
A phase transition associated to the spontaneous breaking of a $U(1)_{L'}$ symmetry, 
where a Majorana mass term is generated, has been 
previously discussed \cite{DiBari:2021dri} as a potential origin for the NANOGrav signal from 12.5-year data set 
\cite{NANOGrav:2020bcs,NANOGrav:2021flc}.
In this case a complex scalar field gets a non-vanishing vacuum expectation value 
at the end of the phase transition and a right-handed neutrino, typically the lightest, coupling to it acquires a Majorana mass. The phase transition involves
only a few additional degrees of freedom forming a dark sector, and some of them can decay into ordinary neutrinos potentially producing extra radiation so that cosmological constraints need to be considered.  It has been shown that these can be respected if the phase transition occurs after
neutrino decoupling and if the dark sector (re-)thermalises only with decoupled ordinary neutrinos. In this case the amount of extra radiation  does not exceed upper bounds from big bang nucleosynthesis and CMB temperature anisotropies. 
However, in \cite{DiBari:2021dri} it was concluded that the amplitude of the NANOGrav signal was too high to be explained by such a phase transition since the
peak of the predicted spectrum was two orders of magnitude below the signal. This conclusion was based on the 12.5-year data, 
and on a way to calculate the sound wave contribution to the GW spectrum valid for values of the strength 
of the phase transition $\alpha \lesssim 0.1$ that is now outdated \cite{Caprini:2015zlo}. 
In this paper we reexamine this conclusion in the light of the 15 year data set and adopting an improved description  of the sound wave contribution, applicable for larger values $\alpha \leq 0.6$ \cite{Cutting:2019zws}.  We introduce
different improvements in the description of the phase transition in the dark sector coupled with neutrinos, 
distinguishing the different temperature and strength parameter compared to the visible sector,  calculating the
ultra-relativistic degrees of freedom at the phase transition occurring during electron-positron annihilations,
including a suppression factor taking into account that  the duration of gravitational wave production
is in general shorter than the duration of the phase transition. 
We confirm that such a phase transition can hardly reproduce 
the whole signal but can be combined with the contribution from the SMBHB baseline model to improve the fit of the signal. 
On the other hand, we notice that the split majoron model receives independent motivations, since it can address different cosmological tensions.
Not only it can ameliorate the well known Hubble tension, and more generally it improves the fit of cosmological observations
compared to the $\Lambda$CDM model, but we notice that it also provides a solution to a potential deuterium problem that is suggested by latest measurement and reanalysis of relevant nuclear rates (D(d,n)$^3$He and D(d,p)$^3$H). 

The paper is structured as follows. In Section 2 we discuss the split majoron model. In Section 3 we discuss the cosmological constraints deriving by the presence of extra-radiation in the model. We also discuss how the model can address 
a potential deuterium problem. In Section 4 we review the calculation of the
GW spectrum and show the results we obtain confronting the NANOGrav 15 year-data set signal. 
In Section 5 we draw our conclusions and discuss future developments . 

\section{The split majoron model}

The split majoron model was sketched in \cite{DiBari:2021dri}. It can be regarded as
an extension at low energies of the multiple majoron model proposed in \cite{DiBari:2023mwu}, 
albeit with important distinctions and phenomenological implications. 
Compared to the traditional majoron model \cite{Chikashige:1980ui}, we have two complex scalar fields
each undergoing its own first order phase transition, one at high scale, above the electroweak scale, and one at much lower
scale, dictated by the possibility to address the NANOGrav signal. If we 
denote by $\phi$ and $\phi'$ the two complex scalar fields, respectively, we can write the Lagrangian as
($I=1,\dots,N$ and $I'=N+1,\dots,N+N'$):
\bea\label{eq:L_l}
-{\cal L}_ {N_I+N_{I'}+\phi+\phi'} & = & 
 \overline{L_{\a}}\,h_{\a I}\, N_{I}\, \widetilde{\Phi} 
+  {\lambda_{I}\over 2}  \, \phi \, \overline{N_{I}^c} \, N_{I} \\  \nonumber
& & +  \overline{L_{\a}}\,h_{\a I'}\, N_{I'}\, \widetilde{\Phi} 
+  {\lambda_{I'}\over 2}  \, \phi' \, \overline{N_{I'}^{c}} \, N_{I'} + V_0(\phi,\phi')
+ {\rm h.c.} \,  ,
\eea
where $\Phi$ is the SM Higgs doublet, $\widetilde{\Phi}$ its dual and the $N_{I}, N_{I'}$ are the RH neutrinos coupling, respectively, to $\phi$ and $\phi'$. Imposing that the Lagrangian (\ref{eq:L_l}) obeys 
a $U(1)_{\sum_I L_I} \times U(1)_{\sum_{I'} L_{I'}}$
symmetry, we can take as (renormalisable) tree level potential (with no $\phi-\Phi$ and $\phi'-\Phi$ couplings)
\be
	V_0(\phi, \phi') = -{\mu^2}\, |\phi|^2 + \lambda \, |\phi|^4 -{\mu'}^{2}|\phi'|^2 + \lambda' |\phi'|^4  
	+ \zeta |\phi|^2 \, |\phi'|^2 \, . \label{Vzero2}
\ee
We will assume that $\phi$ undergoes a phase transition, breaking a $U(1)_{\sum_{I=1}^N L_I}$
global symmetry, at some scale above the electroweak scale. 
In the broken phase we can rewrite $\phi$ as
\be\label{sigma}
\phi ={ e^{i\theta} \over \sqrt{2}}\,\left(v_0 +  S + i \, J \right) \,  ,
\ee 
where $v_0$ is the $\phi$ vacuum expectation value, 
$S$ is a massive boson field with mass $m_S= \sqrt{2 \lambda}\, v_0$ and $J$ is a majoron, a massless Goldstone field.  
The vacuum expectation value of $\phi$ generates RH neutrino masses $M_I = \lambda_I \, v_0/\sqrt{2}$.
After electroweak symmetry breaking, the vacuum expectation value of the Higgs generates 
Dirac neutrino masses $m_{D\a I} = v_{\rm ew}\, h_{\a I} /\sqrt{2}$ and $m_{D\a I'} = v_{\rm ew}\, h_{\a I'} /\sqrt{2}$,
where $v_{\rm ew} = 246\,{\rm GeV}$ is the standard Higgs vacuum expectation value. 
In the case of the RH neutrinos $N_I$, their Majorana masses lead,
via type-I seesaw mechanism, to a light neutrino mass matrix given by the seesaw formula
\begin{eqnarray} \label{eq:seesaw}
(m_\nu)_{\alpha\beta} = - {v_{\rm ew}^2 \over 2} \, {h_{\alpha I} h_{\beta I} \over M_I} \,  .
\end{eqnarray}
Notice that we are writing the neutrino Yukawa matrices in the flavour basis where both charged leptons and 
Majorana mass matrices are diagonal.
The Yukawa couplings $h_{\a I'}$ have to be taken much smaller than usual massive fermions Yukawa 
couplings or even vanishing, as we will point out. 

Eventually, at a scale much below the electroweak scale, $\phi'$ also undergoes a first order phase transition
breaking the $U(1)_{\sum_{I'= 1}^{N'} L_{I'}}$ symmetry. In the broken phase we can rewrite $\phi'$ as
\be\label{sigma}
\phi' ={ e^{i\theta} \over \sqrt{2}}\,\left(v'_0 +  S' + i \, J' \right) \,  ,
\ee 
where $v'_0$ is the $\phi'$ vacuum expectation value, $S'$ is a massive boson field with mass 
$m_{S'}= \sqrt{2 \lambda'}\, v'_0$ and $J'$ is a (second) massless majoron. 
The vacuum expectation value of $\phi'$ generates RH neutrino masses 
$M_{I'} = \lambda_{I'} \, v'_0/\sqrt{2}$. In the following, for the description of the phase transition, 
it will also prove convenient to introduce the real scalar field $\varphi'$, such that $\phi'= (\varphi'/\sqrt{2})\,e^{i\,\theta'}$.

Let us now discuss two different cases we will consider.  
First, we can have a minimal case with $N=2$ and $N'=1$. 
The seesaw formula generates the atmospheric and 
solar neutrino mass scales while the lightest neutrino would be  massless. 
However, after the electroweak symmetry breaking and before the $\phi'$ phase transition, 
the small Yukawa couplings $h_{\a 3}$ generate  a small Dirac neutrino mass for the lightest 
neutrino in a way to have a hybrid case where two neutrino mass 
eigenstates are Majorana neutrinos and the lightest is a Dirac neutrino. 
Finally, at the $\phi'$ phase transition a Majorana mass $M_3$ is generated and 
one has  a second low scale seesaw mechanism (`mini-seesaw') 
giving rise to a lightest neutrino mass $m_1 = \sum_\a |m_{D \a 3}|^2/M_3$.\footnote{Notice
that with our notation $N_3$ is the lightest RH neutrino, not the heaviest.} 

In a second case one has $N=3$ and a generic $N'$. In this case the Yukawa couplings $h_{\a I'}$ can 
even vanish. The RH neutrinos $N_{I'}$ acquire a Majorana mass at the $\phi'$ phase transition but they
do not contribute to the ordinary neutrino masses.  They can be regarded as massive neutral leptons in the dark 
sector, with no interactions with the visible sector (including the seesaw neutrinos). 

As we will better explain in Section 4, the mixing between the two 
complex scalar fields $\phi$ and $\phi'$ significantly increases the strength of the $\phi'$ phase transition.
This will be crucial in enhancing the amplitude of the generated GW spectrum observable in the NANOGrav frequencies. Before delving into the details of the GW production, we discuss some cosmological constraints on our setup and its role in potentially alleviating cosmological tensions.

\section{Cosmological constraints and connection to cosmological tensions}

Let us now consider the impact of cosmological constraints coming
from big bang nucleosynthesis and CMB anisotropies on the model from the amount of extra-radiation 
(also sometimes referred to as {\em dark radiation}). To this end, we first carefully calculate the evolution of the number of degrees of freedom in the model. 

\subsection{Evolution of the ultra-relativistic degrees of freedom in the SM and dark sector}
The number of energy density ultra-relativistic degrees of freedom  $g_{\rho}(T)$ is defined as usual by
$\rho_R(T) \equiv g_{\rho}(T) (\pi^2 /30)\,T^4$, where $\rho_R(T)$ is the energy density in radiation. 
In our case it receives contributions from the SM sector and from the dark sector, so that we can write 
$g_{\rho}(T) = g_{\rho}^{\rm SM}(T) + g_{\rho}^{\rm D}(T)$.  At the $\phi$ phase transition, occurring at 
a phase transition temperature $T_\star$ above the electroweak scale, one has for the SM contribution
$g_{\rho}^{\rm SM}(T_\star) =106.75$ and for the dark sector contribution
\be
g_{\rho}^{\rm D}(T_\star) =  g_{J+S} + {7 \over 4} \, N  \,  ,
\ee
where $g_{J+S} = 2$. Notice here we are assuming that the $N$ seesaw neutrinos all thermalise at 
the $\phi$ phase transition.\footnote{Whether the $N'$ also thermalise, and with them also $\phi'$ or not at high scale, 
it is a question that can answered
only specifying their nature. However, this is not essential, since the $N'$ RH neutrinos can be assumed to decay together
with the $N$ seesaw neutrinos and $J'$ contribution to dark radiation would be anyway very small as the $J$ contribution. 
The important thing is that in any case they thermalise (or rethermalise) prior 
to the low scale phase transition. For definiteness, we assume that $\phi'$ and the $N'$ RH neutrinos
only thermalise prior to the low scale phase transition.} 
This is something that can always be realised since all their decay parameters, defined as
$K_I \equiv (h^\dagger \, h)_{II} \, \bar{v}_{\rm ew}^2 /(M_I \, m_{\rm eq})$ with the effective equilibrium neutrino mass
$m_{\rm eq} = [16\pi^{5/2}\sqrt{g^\star_\rho}/(3\sqrt{5})]\,(v_{\rm eq}/M_{\rm P})$ and 
$\bar{v}_{\rm ew} = v_{\rm ew}/\sqrt{2} = 174\,{\rm GeV}$, 
can be larger than unity in agreement with neutrino oscillation experiments. 
Therefore, at the high scale phase transition the dark sector
is in thermal equilibrium with the SM sector thanks to the seesaw neutrino Yukawa couplings. 

After the $\phi$ phase transition, all massive particles in the dark sector, $S$ plus the $N$ seesaw neutrinos will quickly decay,  while the massless majoron $J$ will contribute to dark radiation. 
We can then track the evolution of $g_{\rho}(T)$ at temperatures below $T_{\star}$
and prior to the low scale phase transition occurring at a temperature $T_{\star}'$
and also prior to any potential process of {\em rethermalisation} of the dark sector
that we will discuss later.  

In particular, we can focus on temperatures $T \ll m_\mu \sim 100\,{\rm MeV}$.
In this case the SM contribution\footnote{For our purposes it is certainly sufficient to treat neutrinos as fully thermal, neglecting 
the small non-thermal contribution produced by $e^+$ -- $e^-$ annihilations. However, we will take into account
this small contribution in the calculation of the amount of extra-radiation.}  can be written as \cite{DiBari:2018vba}
\be
g^{\rm SM}_{\rho}(T \ll 100\,{\rm MeV}) = 
g^{\g+e^{\pm}+3\nu}_{\rho}(T) = 
2 + {7\over 8}\,\left[4\, g_{\rho}^{e}(T) + 6\,  r_\nu^4(T)  \right]\,  ,
\ee 
where the number of energy density ultra-relativistic  degrees of freedom of electrons per single spin degree is given by
\be
g_\rho^e(T) ={120\over 7\, \pi^4}\, \int_0^\infty \, dx \, {x^2 \,\sqrt{x^2 + z^2} \over e^{\sqrt{x^2 + z^2}} + 1} \,  ,
\ee
with $z \equiv m_e/T$.  Above the electron mass one has $g_\rho^e(T \gg m_e) = 1$, while  of course $g_\rho^e(T ) \ra 0$ 
for $T / m_e \ra 0$.  
The neutrino-to-photon temperature ratio $r_{\nu}(T) \equiv T_{\nu}(T)/T$ can, as usual, be
calculated using entropy conservation,
\be\label{rnuT}
r_{\nu}(T)  = \left({2\over 11}\right)^{1\over 3} \, \left[g_s^{\gamma + e^{\pm}}(T)\right]^{1\over 3} \,  ,
\ee 
where 
\be
g_s^{\gamma + e^{\pm}}(T) = 2 + {7\over 2}\,g^e_s(T)\,  ,
\ee
having defined the contribution to the number of entropy density ultra-relativistic  
degrees of freedom of electrons (per single spin degree of freedom) as
\be
g_s^e(T) = {8\over 7}{45 \over 4\pi^4}\, \int_0^\infty \, dx \, {x^2 \,\sqrt{x^2 + z^2} +{1\over 3}\,{x^4 \over \sqrt{x^2 + z^2}} \over e^{\sqrt{x^2 + z^2}} + 1} \,   .
\ee
One can again verify that $g_s^e(T \gg m_e) = 1$ and $g_s^e(T) \ra 0 $ for $T / m_e \ra 0$,
so that one recovers the well known result $r_\nu (T\ll m_e) = (4/11)^{1/3}$.  With this function one can write the SM 
number of entropy density ultra-relativistic  degrees of freedom as
\be
g^{\rm SM}_s(T \ll m_\mu ) = g^{\g+e^{\pm}+3\nu}_{s}(T) = 2 + {7\over 8}\,\left[4 \, g_{s}^{e}(T) + 6\, r^3_{\nu}(T)  \right]\,   .
\ee 
For $T \ll m_e$ one recovers the well known results $g^{\rm SM}_s(T \ll m_e ) = 43/ 11  \simeq 3.91 $
and $g^{\rm SM}_\rho(T \ll m_e ) \simeq 3.36 $.\footnote{If one includes the small non-thermal
neutrino contribution, these numbers are corrected to $3.93$ and $3.385$, respectively. As we said, at this stage,
this small correction can be safely neglected.}

Let us now focus on the dark sector contribution. This is very easy to calculate since one has simply
$g_{\rho}^{\rm D}(T) = g_{J} \, [r_{\rm D}(T)]^4$ and 
$g_{s}^{\rm D}(T) = g_{J} \, [r_{\rm D}(T)]^3$, where $g_J =1$ and where 
the dark sector-to-photon temperature ratio $r_{\rm D}(T)$ 
can again be calculated from entropy conservation as
\be\label{rDsimple}
r_{\rm D}(T) =  \left[ {g^{\rm SM}_s(T) \over g^{\rm SM}_s(T_\star)} \right]^{1/3} \,   .
\ee
For example, for $m_\mu \gg T \gg m_e$, one finds $r_{\rm D}(T) = (43/427)^{1/3} \simeq 0.465$.
We can also rewrite $g_{\rho}^{\rm D}(T)$ in terms of the extra-effective number of neutrino
species $\Delta N_{\nu}(T)$ defined by
\be
g_{\rho}^{\rm D}(T) \equiv {7\over 4}\, \Delta N_\nu (T) \, [r_\nu(T)]^{4} \, . 
\ee
Again, in the particular case $m_\mu \gg T \gg m_e$, one finds 
\be
\Delta N_{\nu}(T) = {4 \over 7} \, g_J \, [r_{\rm D}(T)]^4  = {4\over 7}\,
\left({43 \over  427}\right)^{4\over 3} \simeq 0.027 \, .  
\ee
Such a small amount of extra radiation is in agreement with all cosmological constraints that we summarise here:
\begin{itemize}
\item Primordial helium-4 abundance measurements combined with the baryon abundance extracted from 
cosmic microwave background (CMB) anisotropies place a constraint on $\D N_\nu^{\rm eff}(t)$ at $t=t_{\rm f} \sim 1\,{\rm s}$,
the time of freeze-out of the neutron-to-proton ratio~\cite{Fields:2019pfx}:
\be\label{ubDNeetf}
\D N_\nu(t_{\rm f}) \simeq  -0.1 \pm 0.3 \ \Rightarrow \  \D N_\nu(t_{\rm f}) \lesssim 0.5 \, \ \  (95\% \, {\rm C.L.})\,  .
\ee
\item From measurements of the primordial deuterium abundance 
at the time of nucleosynthesis, $t_{\rm nuc} \simeq 310 \, {\rm s}$, 
corresponding to $T_{\rm nuc} \simeq 65\,{\rm keV}$ \cite{Pisanti:2020efz}:
\be\label{ubDNeetnuc}
\D N_\nu(t_{\rm nuc}) =  -0.05 \pm 0.22 \  \Rightarrow \  \D N_\nu(t_{\rm nuc}) \lesssim 0.4 \, \ \ (95\% \, {\rm C.L.}) \,  .
\ee
\item CMB temperature and polarization anisotropies constrain $\D N_\nu(t)$ at recombination, when $T \simeq T_{\rm rec} \simeq 0.26 \, {\rm eV}$.  The {\em Planck} collaboration finds\footnote{This result is found ignoring the astrophysical
measurement of $H_0$, i.e., the Hubble tension.}~\cite{Aghanim:2018eyx}
\be
\D N_\nu(t_{\rm rec}) =  -0.05 \pm 0.17\  \Rightarrow \  \D N_\nu(t_{\rm rec}) \lesssim 0.3 \, \ \ (95\% \, {\rm C.L.}) \,  .
\ee
\end{itemize}

Let us now consider the low scale phase transition, assuming first that this occurs at a temperature $T'_\star$
above neutrino decoupling temperature $T_{\nu}^{\rm dec} \sim 1\,{\rm MeV}$, 
so that $r_{\nu}(T'_\star) = 1$, but below 1 GeV.  At such low temperatures,  
Yukawa couplings are ineffective to rethermalise the dark sector \cite{DiBari:2021dri}. 
On the other hand, the coupling term $\zeta \, J^2 \, |\phi'|^2 $
can thermalise $\phi'$ and the $N'$ RH neutrinos with $J$ at a common temperature $T_{\rm D}$.
Therefore, at the $\phi'$ phase transition, the dark sector will have a temperature $T'_{{\rm D}\star} \simeq 0.465 \, T'_{\star}$ 
and with such a small temperature one would obtain a GW production much below the NANOGrav signal. Notice that
after the phase transition the second  majoron $J'$ would give a contribution to $\Delta N_{\nu}(T)$ 
equal to that one from $J$, in a way that one would obtain $\Delta N_{\nu}(T) \simeq 0.05$. 

One could envisage some interaction able to rethermalise the dark sector so that $r_{\rm D}(T'_{\star}) = 1$.
However, in this case the thermalised $J'$ abundance would correspond to have $\Delta N_{\nu}(T) \simeq 8/14 \simeq 0.6$ throughout BBN and recombination, in disagreement with the cosmological constraints we have just reviewed.\footnote{A possible
interesting caveat to this conclusion is to modify the model introducing an explicit symmetry breaking term that would
give $J'$ a mass. In this way $J'$ might decay prior to neutron-to-proton ratio freeze out, thus circumventing all constraints. We will be back in the final remarks on realising this scenario that, in the context of a general phase transition
in a dark sector, has been discussed in \cite{Fairbairn:2019xog,Bringmann:2023opz}.} For this reason we now consider, as in  \cite{DiBari:2021dri},
the case when the low scale phase transition 
occurs well below neutrino decoupling (i.e., $T'_\star \lesssim 1\,{\rm MeV}$).

In this case a rethermalisation between the dark sector and just the decoupled ordinary neutrino background
can occur without violating the cosmological constraints. 
Prior to the $\phi'$ phase transition one has the interactions\footnote{The couplings
$\widetilde{\lambda}_i$ can be related to the couplings $\lambda_I$ in Eq.~(\ref{eq:L_l}).}
\be\label{nudark}
-{\cal L}_{\nu{\rm D}} = {i\over 2}\,\sum_{i=2,3}\widetilde{\lambda}_i \, \overline{\nu_{i}} \, \g^5 \,\nu_{i} \, J
\ee
that can thermalise the majoron $J$ with the ordinary neutrinos, and also the complex scalar field $\phi'$ via the coupling $\zeta\, J^2 |\phi'|^2$. This interacts with  the  $N_{I'}$'s 
that also thermalise prior to the phase transition. The lightest neutrino $\nu_1$ thermalises after the $\phi'$ 
phase transition interacting with $J'$ via an interaction term analogous to the one in Eq.~(\ref{nudark}).
In this way ordinary neutrinos would lose part of their energy
that is transferred to the dark sector, so that they reach a common temperature $T_{\nu{\rm D}}$
given by\footnote{This expression assumes that the initial temperature of the
dark sector is vanishing. However, the majoron $J$ was thermalised at the $\phi$-phase transition
and afterwards its temperature is described by Eq.~(\ref{rDsimple}). 
If this small initial temperature is taken into account, then Eq.~(\ref{TnuD}) 
gets generalised into 
\be\label{rnuDgen}
 r_{\nu{\rm D}} = \left({N_{\nu}^{\rm SM}(T) r^4_{\nu}(T) + 
 (4/7) r^4_D(T)\, \over N_{\nu}^{\rm SM}(T) + N' + 12/7 + 4\,\D g/7} \right)^{1 \over 4} \,  .
\ee
On the other hand, the correction is quite small and we can safely neglect it.
} \cite{Chacko:2003dt,Escudero:2019gvw,Escudero:2021rfi} 
\be\label{TnuD}
 r_{\nu{\rm D}} \equiv {T_{\nu{\rm D}}\over T} = r_{\nu}(T) \, \left({N_{\nu}^{\rm SM}(T) \over N_{\nu}^{\rm SM}(T) + N' + 12/7 + 4\,\D g/7} \right)^{1 \over 4} \,  ,
\ee
where $T_{\nu}(T)$, given by Eq.~(\ref{rnuT}), is the standard neutrino temperature (i.e., in the absence of the dark sector)
so that one simply has $r_{\nu {\rm D}}(T) = r_{\nu}(T)\, [T_{\nu{\rm D}}(T)/T_{\nu}(T)] $. Notice that $N_{\nu}^{\rm SM}(T \gg m_e/2) = 3$
and $N_{\nu}^{\rm SM}(T \ll m_e/2) = 3.043$ for the predicted SM value of the effective neutrino species \cite{Cielo:2023bqp}.    

\subsection{Hubble tension}
The minimal content of the dark sector is given by $J,\phi'$ and $N'$ RH neutrinos. However, we can also account for
the possibility of the existence of $\D g$ extra massless degrees of freedom. 
For example, for $N' =1 $ and $\D g =0, 1, 2, 3$, one finds respectively $T_{\nu{\rm D}}/T_\nu = 0.815, 0.784, 0.76$.
One can also calculate the amount of extra-radiation at temperatures much below the electron mass, obtaining
\be\label{extra}
\Delta N_\nu \simeq 3.043 \left[\, \left({3.043 + N' + 12/7 +4\Delta g /7 \over 3.043 + N' + 12/7 + 4\Delta g /7  - N_{\rm h}} \right)^{1\over 3} - 1 \right]  \,  ,
\ee
where $N_{\rm h}$ is the number of massive states that decay after the phase transition and produce the excess radiation.\footnote{The expression Eq.~(\ref{extra}) is obtained assuming entropy conservation and neglecting the initial 
small amount of majorons $J$ from the first thermalisation of the dark sector at high scale. 
If this is taken into account, then Eq.~(\ref{extra}) gets generalised into
\be\label{extragen}
\Delta N_\nu \simeq 3.043 \left[\, \left({3.043 + N' + 12/7 +4\Delta g /7 \over 3.043 + N' + 12/7 + 4\Delta g /7  - N_{\rm h}} \right)^{1\over 3}\left(1 + {r^4_D \over r^4_\nu} \right) - 1 \right]  \,  ,
\ee
where $r_D$ is given by Eq.~(\ref{rDsimple}). This more general expression
might be useful in the case of a higher number of decoupled degrees of freedom in the dark sector in addition
to $J$. For example, if one considers the case of multiple majorons giving mass to the seesaw neutrinos, as considered in \cite{DiBari:2023mwu}.
}
In our case these states are given by $S'$ and the $N'$ RH neutrinos so that $N_h = N' + 1$. For $N' =1$ and $\D g =0$ one obtains $\Delta N_\nu \simeq 0.465$. In this case, such an amount of extra-radiation can actually even be beneficial in order to ameliorate the Hubble tension \cite{Escudero:2019gvw,Blinov:2020hmc} compared to the $\Lambda$CDM model since one has a simultaneous injection of extra radiation
together with a reduction of the neutrino free streaming length due to the interactions between the ordinary neutrinos
and the majorons. For this reason a low energy scale majoron model of this kind is a leading candidate to resolve the cosmological tensions
within the $\Lambda$CDM model \cite{Schoneberg:2021qvd}.
Recently a new analysis of this model has been presented in \cite{Sandner:2023ptm} where
the authors find an improvement at the level of $1 \s$ compared to the $\L$CDM model. It is then interesting 
that this kind of model can link the NANOGrav signal, that we are going to discuss in the next section, to the cosmological tensions.  

\subsection{Deuterium problem}
If the rethermalisation occurs at a temperature above 65 keV, one should worry  about the constraint Eq.~(\ref{ubDNeetnuc}) from deuterium. In this case, one can reduce the amount of
extra-radiation increasing the number of massless degrees of freedom in the dark sector considering $\Delta g \neq 0$.
For example, for $\Delta g = 1,2,3$ one obtains, respectively, $\Delta N_\nu = 0.41, 0.37, 0.33$. 
Therefore, an increase of the degrees of freedom in the dark sector actually produces a reduction of the amount of extra-radiation making it compatible also with deuterium constraints. However, notice that it is actually interesting that the model 
predicts some increase of the deuterium abundance compared to standard big bang nucleosynthesis (SBBN). There is indeed a potential tension with the current measurement of primordial deuterium abundance within SBBN.  The experimental value is found to be \cite{Cooke:2017cwo} ${\rm D/H} = (2.527 \pm 0.030) \times 10^{-5}$. Using a calculation of D(d,n)$^3$He and 
D(d,p)$^3$H nuclear rates based on
theoretical ab-initio energy dependencies  the authors of \cite{Pitrou:2020etk} find, as SBBN prediction,
${\rm D/H} = (2.439 \pm 0.037) \times 10^{-5}$, showing a $\sim 2\sigma$ tension with the experimental value.
Since the primordial  deuterium abundance scales with $\Delta N_\nu$ 
approximately as \cite{DiBari:2001ua} $({\rm D/H})(\Delta N_\nu) = ({\rm D/H})^{\rm SBBN} (1 + 0.135\, \Delta N_\nu)^{0.8}$, one finds that $\Delta N_{\nu}(t_{\rm nuc}) \simeq 0.3$ would solve the tension. 
However, using a polynomial expansion of the S-factors of the above-mentioned nuclear rates
the authors of \cite{Pisanti:2020efz} find ${\rm D/H} = (2.54 \pm 0.07) \times 10^{-5}$, a predicted value 
that would be essentially in agreement with the experimental value and that places the upper bound on $\Delta N_{\nu}(t_{\rm nuc})$
given in Eq.~(\ref{ubDNeetnuc}).  New and more accurate data on the nuclear rates should be able to establish
which one of the two descriptions is more reliable, thus confirming or ruling out the tension \cite{Pitrou:2021vqr}. 
In case it will be confirmed, the split majoron model would be not only a natural candidate to explain the tension but, 
very importantly,  it would also offer a simultaneous solution to the other cosmological tensions and, as we are now going to discuss, 
realise an intriguing connection with the NANOGrav signal.

\section{GW spectrum predictions confronting the NANOGrav signal}
We first briefly review how the first order phase transition parameters relevant for the production of GW spectrum in the split majoron model are calculated and refer the interested reader to Ref.~\cite{DiBari:2023mwu} for a broader discussion.\footnote{\textcolor{black}{Production of GWs from first order phase transitions in the dark sector has been discussed, in a general framework,
in \cite{Breitbach:2018ddu,Fairbairn:2019xog,Caprini:2019egz,Bringmann:2023opz}}.}
The finite-temperature effective potential for the (real) scalar $\varphi'$ can be calculated perturbatively at one-loop level and is the summation of zero temperature tree-level, one-loop Coleman-Weinberg potential and one-loop thermal potential. Using thermal expansion of the one-loop thermal potential, this can be converted in a \emph{dressed effective potential} given by
\be\label{VTeffminimal2}
 V^{T_{\nu{\rm D}}}_{\rm eff}(\varphi') \simeq {1\over 2}\, \widetilde{M}_{T_{\nu{\rm D}}}^2\,\varphi^{'2} - 
 (A \, T_{\nu{\rm D}}+C) \, \varphi^{'3} + \frac{1}{4}\lambda_{T_{\nu{\rm D}}}\, \varphi^{'4} \,  ,
\ee
where notice that the common {\em neutrino-dark sector temperature} $T_{\nu{\rm D}}$ replaces the photon temperature $T$.
However, once the calculations are done in terms of $T_{\nu{\rm D}}$, everything can then be more conveniently
expressed in terms of the standard $T$ simply using $T_{\nu{\rm D}}(T) = r_{\nu{\rm D}}(T)\,T$.
Here, a zero-temperature cubic term $C = \zeta^2 v_0'/(2\lambda)$ is introduced due to the presence of the scalar $\phi$ with a high scale vacuum expectation value during the phase transition of $\phi'$ at a lower scale. This term significantly enhances the strength of the phase transition. The other parameters in Eq.~\eqref{VTeffminimal2} are given by
\be\label{MTsq}
\widetilde{M}_{T_{\nu{\rm D}}}^2\equiv 2\,D\,(T_{\nu{\rm D}}^2 - \overline{T}_{\nu{\rm D}}^2) \,   ,
\ee
where the destabilisation temperature $\overline{T}_{\nu{\rm D}}$ is  given by
\be\label{DT0sq}
2\,D\,\overline{T}_{\nu{\rm D}}^2 =   \lambda'\,v_0^{'2} +{N' \over 8\,\pi^2}\,{M^{'4} \over v_0^{'2}} 
-{3\over 8 \,\pi^2}\lambda^{'2} \, v_0^{'2}  \,   ,
\ee
and the dimensionless constant coefficients $D$ and $A$ are expressed as
\be\label{DA}
D = {{\lambda' \over 8} + {N'\over 24}\,{M^{'2} \over v_0^{'2}}} \,   \;\;\;\; \mbox{\rm and} \;\;\;\;
A =  {(3\,\lambda')^{3/2} \over 12\pi } \,  \,  .
\ee
The dimensionless temperature dependent quartic coefficient $\lambda_{T_{\nu{\rm D}}}$ is given by
\be\label{lambdaT}
\lambda_{T_{\nu{\rm D}}} = \lambda'  - \frac{N'\, M^{'4}}{8\,\pi^2 \, v_0^{'4}}  \, \log {a_F \, T_{\nu{\rm D}}^2 \over e^{3/2}\,M^{'2}} 
+  {9\lambda^2 \over 16 \pi^2}\,\log {a_B \, T_{\nu{\rm D}}^2 \over e^{3/2}\,m_S^{'2}} \,  .
\ee
The cubic term is negligible at very high temperatures and the potential is symmetric with respect to $\phi'$. 
However, at lower temperatures it becomes important and a stable second minimum forms at a nonzero $\phi'$.
At the critical temperature the two minima are degenerate and below the critical temperature 
bubbles can nucleate from the false vacuum to the true vacuum with nonzero probability. We refer to 
$T'_{{\nu{\rm D}}\star}$ as the characteristic phase transition temperature and identify it with the {\em percolation temperature}, 
when $1/e$ fraction of space is still in the  false vacuum. It is related to the
corresponding temperature of the SM sector simply by $T'_{{\nu{\rm D}}\star}= T'_\star \, r_{\nu{\rm D}}(T'_\star)$
(we always use $T$ as independent variable and from this we calculate $T_{\nu{\rm D}}$). 
Two other parameters relevant for the calculation of the GW spectrum from first order phase transitions are $\alpha$ and $\beta/H_\star$, where the first denotes the strength of the phase transition and the latter describes the inverse of the duration of the phase transition.  These parameters are defined as
\be\label{betaoverH}
{\beta \over H_{\star}} \simeq T'_{\star} \left.{d(S_3/T_{\nu{\rm D}}) \over dT}\right|_{T'_{\star}} \,  , \qquad \text{and}\qquad \a \equiv {\ve(T'_{{\nu{\rm D}}{\star}}) \over \rho(T'_{\star})} \,  ,
\ee
where $S_3$ is the spatial Euclidean action, $\ve(T'_{{\nu{\rm D}}\star})$ is the latent heat released during the phase transition and $\rho(T'_{\star})$ is the
total energy density of the plasma, including both SM and dark sector degrees of freedom. An approximate analytical estimate for calculating $S_3/T_{\nu{\rm D}}$, and from this $T'_{{\nu{\rm D}}\star}$, in terms of the model parameters can be found in Ref.~\cite{DiBari:2023mwu}. In calculating $\alpha$ for phase transition at low temperatures, one must be careful about various cosmological constraints, as outlined in section 3. 

We now proceed to calculate the GW spectrum of the model relevant for nanoHZ frequencies. Assuming that first order phase transition occurs in the detonation regime where bubble wall velocity $v_w > c_s = 1/\sqrt{3}$, the dominant contribution to the GW spectrum mainly comes from sound waves in the plasma. 
Numerical simulations confirm for $\alpha \lesssim 0.1$  
\cite{Caprini:2015zlo,Hindmarsh:2017gnf,Weir:2017wfa} the analytical result from the sound shell model\footnote{\textcolor{black}{It assumes that the sound waves are linear and that their power spectrum is determined by the characteristic form of the sound shell around the expanding bubble.}} \cite{Hindmarsh:2016lnk}:
\be\label{omegasw}
h^2\Omega_{\rm sw 0}(f) =3 \, h^2 \, r_{\rm gw}(t_\star,t_0)\,\widetilde{\Omega}_{\rm gw} \,  
H_\star \, R_\star \, \left[\frac{\kappa(\a_{\nu{\rm D}})\, \alpha}{1+\alpha}\right]^2 \, 
\widetilde{S}_{\rm sw} (f) \, \Upsilon(\a,\alpha_{\nu{\rm D}},\beta/H_{\star})\, ,
\ee
where $R_\star$ is the mean bubble separation and a standard relation is 
$R_\star = (8\pi)^{1/3}\,v_{\rm w}/\beta$. 
Notice that the parameter 
$\alpha_{\nu{\rm D}} \equiv \ve(T'_{\star})/\rho_{\nu{\rm D}}(T'_\star)$ replaces $\a$ inside $\kappa$ and $\Upsilon$
\cite{Fairbairn:2019xog,Bringmann:2023opz}, where we simple defined $\rho_{\nu{\rm D}}\equiv \rho_\nu + \rho_{\rm D}$.
We adopt Jouguet detonation solution for which the efficiency factor is given by ~\cite{Steinhardt:1981ct,Espinosa:2010hh} 
\begin{eqnarray}
\kappa(\a_{\nu{\rm D}}) \simeq {\alpha_{\nu{\rm D}}\over 0.73+0.083\sqrt{\alpha_{\nu{\rm D}}}+\a_{\nu{\rm D}}} \,,
\end{eqnarray}
and the bubble wall velocity $v_{\rm w}(\a_{\rm D}) = v_{\rm J}(\a_{\rm D})$, where
\begin{eqnarray} \label{eq:Jouguet}
v_{\rm J}(\a_{\nu{\rm D}}) \equiv \frac{\sqrt{1/3} + \sqrt{\alpha_{\nu{\rm D}}^2 +2\alpha_{\nu{\rm D}}/3}}{1+\alpha_{\nu{\rm D}}}\,.
\end{eqnarray}
The suppression factor $\Upsilon(\alpha,\alpha_{\nu{\rm D}},\beta/H_\star) \leq 1$ takes into account the finite 
lifetime of the soundwaves
and is given by
\cite{Ellis:2018mja,Guo:2020grp}:
\be
\Upsilon(\alpha,\alpha_{\nu{\rm D}},\beta/H_\star) =  1- {1\over \sqrt{1+ 2\, H_{\star} \tau_{\rm sw}}} \,  ,
\ee
where we can write 
\be
H_\star \tau_{\rm sw}  =  (8\,\pi)^{1\over 3}{v_{\rm w}\over \beta/H_{\star}} \left[ {1 + \a \over \kappa(\a_{\nu{\rm D}}) \, \alpha}\right]^{1/2} \,  .
\ee
Only in the ideal asymptotic limit $H_{\star} \tau_{\rm sw} \ra \infty$ one has $\Upsilon = 1$. 
The prefactor $\widetilde{\Omega}_{\rm gw}$ is a dimensionless number given by an integral
over all wave numbers $k$ \cite{Hindmarsh:2015qta}
\be
\widetilde{\Omega}_{\rm gw} = \int {dk \over k} \, {(k\,L_{\rm f})^3 \over 2\pi^2} \, \widetilde{P}_{\rm GW}(k\,L_{\rm f}) \ ,
\ee 
where $L_{\rm f}$ is a characteristic length scale 
in the velocity field, $\widetilde{P}_{\rm GW}$ is the GW power spectrum 
and it is found \cite{Hindmarsh:2015qta,Hindmarsh:2017gnf}
\be
\widetilde{\Omega}_{\rm gw} = {(0.8 \pm 0.1)  \over 2 \pi^3} \sim 10^{-2} \,  .
\ee 
The redshift factor $r_{\rm gw}(t_\star,t_0) $, evolving $\Omega_{\rm gw\star} \equiv \rho_{\rm gw\star}/\rho_{{\rm c}\star}$ to $\Omega_{\rm gw 0} \equiv \rho_{\rm gw 0}/\rho_{\rm c 0}$, is given by 
\cite{Kamionkowski:1993fg}
\be
r_{\rm gw}(t_\star,t_0) = \left({a_\star \over a_0}\right)^4 \, \left({H_\star \over H_0}\right)^2 
= \left({g_{s0}\over g_{s\star}}\right)^{4 \over 3}\,{g_{\r\star} \over g_\gamma} \, \O_{\gamma 0} \,  ,
\ee
where $\Omega_{\gamma 0} \equiv \rho_{\gamma 0}/\rho_{{\rm c} 0}$.
The normalised spectral shape function $\widetilde{S}_{\rm sw} (f)$ is given by 
$\widetilde{S}_{\rm sw} (f) \simeq 0.687\, S_{\rm sw} (f) $ with
\begin{eqnarray}\label{Ssw}
S_{\rm sw} (f) = \left(\frac{f}{f_{\rm sw}}\right)^3 \left[\frac{7}{4+3({f/f_{\rm sw}})^2} \right]^{7/2} \,  ,
\end{eqnarray} 
where $f_{{\rm sw}}$ is the peak frequency at the present time. This is simply obtained redshifting the peak frequency 
at the time of the phase transition: $f_{{\rm sw}} = f_{{\rm sw}\star} a_{\star}$. The
peak frequency at the phase transition is given, in terms of $v_{\rm w}$ and $\beta/H_\star$, by  \cite{Hindmarsh:2017gnf}
\be
f_{{\rm sw}\star} = \kappa \, {\beta/H_\star \over v_{\rm w}} \, H_{\star} \,  ,
\ee
with $\kappa \simeq 0.54$.
From entropy conservation one can write $a_\star = T_0\,g_{\rm s}^{1/3}(T\ll m_e)/(T_\star\,g^{1/3}_{\rm s\star})$ and 
from the Friedmann equation $H_\star \simeq 1.66\, T_\star^2\, g_{\rho \star}^{1/2}/M_{\rm Pl} $, where 
$T_0 \simeq 2.35 \times 10^{-4}\,{\rm eV}$ is the CMB temperature 
and $g_{s}(T\ll m_e) = g_{s}^{\rm SM}(T\ll m_e) \simeq 3.91$. 
In this way one obtains for the peak frequency at the present time
\be\label{fswgen}
f_{{\rm sw}} \simeq 1.66 \, \kappa \, T_0 \, g_{{\rm S}0}^{1/3} \, {\beta/H_\star \over v_{\rm w}}
\,{T_\star \, g_{\rho \star}^{1/2} \over M_{\rm Pl} \, g^{1/3}_{\rm S\star}}
\simeq 
4.1\,\mu{\rm Hz} \, \frac{1}{v_{\rm w}} \frac{\beta}{H_\star}  \frac{T_\star}{\rm 100\,GeV} 
\frac{g_{\rho\star}^{1/2}}{g_{S\star}^{1/3}} 
\ee
For phase transitions above the electroweak scale the peak frequency 
one has $g_{\rho \star} = g_{s \star} = g_\star$. In this way Eq.~(\ref{fswgen}) specialises into 
\begin{eqnarray} \label{fpeak}
f_{\rm sw} =8.9\,\mu{\rm Hz} \, \frac{1}{v_{\rm w}} \frac{\beta}{H_\star}  \frac{T_\star}{\rm 100\,GeV}  
 \, \left( \frac{g_\rho^\star}{106.75} \right)^{1/6} \, .
\end{eqnarray}
We can also write a numerical expression for the redshift factor
\be
r_{\rm gw}(t_\star,t_0) \simeq 3.5 \times 10^{-5} \, 
\left({106.75 \over g_{\star}} \right)^{1\over 3} \,  \left({0.68 \over h}\right)^2 \, 
\ee
and, finally, for the GW spectrum from sound waves\footnote{This numerical expression agrees with the 
one in  \cite{Weir:2017wfa} (see Erratum in v3).}
\be\label{omegasw2}
h^2\Omega_{\rm sw 0}(f) =0.97\times 10^{-6}\, {\widetilde{\Omega}_{\rm gw} \over 10^{-2}} \,
\frac{v_{\rm w}(\a)}{\beta/H_\star} \left[\frac{\kappa(\a_{\nu{\rm D}})\, \alpha}{1+\alpha}\right]^2  
\left( \frac{106.75}{ g_\star} \right)^{1/3}  \, S_{\rm sw} (f) \, \Upsilon(\alpha,\alpha_{\nu{\rm D}},\beta/H_\star)
\,  .
\ee
Let us now turn to the case of our interest, a low scale phase transition for $T'_\star \lesssim 1\, {\rm MeV}$.
In this case one has $g_\rho(T'_\star) \neq g_{s}(T'_\star)$, specifically:
\bea
g_{\rho\star}' \equiv g_\rho(T'_\star) & = & g^{\g+e^{\pm}}_{\rho}(T'_\star) + g^{3\nu}_{\rho}(T'_\star) + g_\rho^{\rm D}(T'_\star)  \\ \nonumber
& = & 2 + {7\over 2}\, g_{\rho}^{e}(T'_{\star}) + \left( {21 \over 4}\,  + 
g_J+g_{S'} + g_{J'}+ {7\over 4} \, N' + \D g \right)\,r_{\nu{\rm D}}^4(T'_\star) \,   ,
\eea
and 
\bea
g_{s\star}' \equiv g_s(T'_\star) & = & g^{\g+e^{\pm}}_{s}(T'_\star) + g^{3\nu}_{s}(T'_\star) + g_\rho^{\rm D}(T'_\star)  \\ \nonumber
& = & 2 + {7\over 2}\, g_{s}^{e}(T'_{\star}) + \left( {21 \over 4} + g_J+g_{S'} + g_{J'}+ {7\over 4} \, N' + \D g \right) \,r_{\nu{\rm D}}^3(T'_\star) \,   .
\eea
In the limit $T'_\star \gg m_e/2$ one has $r_{\nu{\rm D}} \simeq g_{\rho}^e \simeq 1$.
If, for definiteness, we consider the minimal case with $\D g =0$ and $N' = 1$,
one has then, for $T'_\star \gg m_e/2$,  
$g_{\rho\star}'  \simeq g_{s\star}' \simeq 62/4$.
We can then conveniently rewrite numerically:
\be
r_{\rm gw}(t'_\star,t_0) \simeq 6.6 \times 10^{-5} \, \left({0.68 \over h}\right)^2 \,
\left(\frac{15.5}{ g'_{s\star}}\right)^{4\over 3} \, \frac{g'_{\rho\star}}{15.5} \,  ,
\ee
\be
f_{{\rm sw}} \simeq 
6.47\,{\rm nHz} \, \frac{1}{v_{\rm w}} \frac{\beta/H_\star}{100}  \frac{T_\star}{\rm 1\,MeV} 
\frac{(g_{\rho\star}/15.5)^{1/2}}{(g_{s\star}/15.5)^{1/3}} 
\ee
and 
\be
h^2\Omega_{\rm sw 0}(f) =1.845 \times 10^{-6} \, {\widetilde{\Omega}_{\rm gw} \over 10^{-2}} \,
\frac{v_{\rm w}(\a)}{\beta/H_\star} \left[\frac{\kappa(\a_{\nu{\rm D}})\, \alpha}{1+\alpha}\right]^2  
\left( \frac{15.5}{ g'_{s\star}} \right)^{4/3}\left(\frac{g'_{\rho\star}}{15.5}\right)  
\, S_{\rm sw} (f) \, \Upsilon(\alpha,\alpha_{\nu{\rm D}},\beta/H_\star) \,  .
\ee
For $\alpha \gtrsim 0.1$ one expects strong deviation from (\ref{omegasw}) that can be expressed
in terms of a function $\xi(f;\a,v_{\rm w},\beta/H_\star,\dots)$. This function is currently undetermined. 
Here we  mention a few effects that have been studied and that contribute to $\xi(f;\a,v_{\rm w},\beta/H_\star,\dots)$.
\begin{itemize}
\item The expression (\ref{omegasw}) neglects a contribution from turbulent motion of the dark sector plasma after the phase transition.
This contribution is certainly subdominant for $\alpha \lesssim 0.1$ but it might become sizeable for $\alpha \gtrsim 0.1$,
though its determination requires a better theoretical understanding \cite{Auclair:2022jod}.
\item In \cite{Cutting:2019zws}
it was found in numerical simulations that for the integral on the whole range of frequencies, i.e., for the 
sound wave contribution to the GW energy density parameter, one has a  suppression  by a factor $0.1$--$1$
for values $\a \lesssim 0.6$ and $v_{\rm w} \simeq c_{\rm s}$ compared to the expected result one obtains
integrating Eq.~(\ref{omegasw}).  There are currently no well established results 
on the spectral shape function deviation for $\a > 0.1$ compared to the broken power law Eq.~(\ref{Ssw}).
In order to account for such an indetermination, we show the GW spectrum for bands corresponding to $\xi = 0.1$--$1$ in our plots in Figs.~\ref{fig:fig1} and \ref{fig:fig2}.\footnote{Recently, numerical results have been derived showing that 
a steep  $S_{\rm sw}(f) \propto f^7$ growth may appear below the peak under certain circumstances, leading to a bump 
in the spectral shape \cite{RoperPol:2023dzg}. The presence of this potential bump could potentially
lead to a clear signature in NANOGrav data.}
 
 We refer the interested reader to Ref.~\cite{DiBari:2023mwu} for more details about the known issues and caveats in using the above expressions for calculating the GW spectrum. The GW spectrum plots are obtained for a set of benchmark points given in Table \ref{table:BP1} and \ref{table:BP2} in Figs. \ref{fig:fig1} and \ref{fig:fig2}, respectively. 
\item Recently, it has been found in \cite{Ajmi:2022nmq} that,  when the bubbles expand as deflagrations, the heating of the fluid in front of the phase boundary suppresses the nucleation rate increasing the mean bubble separation and enhancing
the gravitational wave signal by a factor of up to order ten. This enhancement increases 
for increasing values of $\alpha$ and low values of $v_{\rm w}$, so that it is sizeable only in the
case of deflagrations ($v_{\rm w} < c_{\rm s}$), while it is negligible in the case of detonations ($v_{\rm w} > c_{\rm s}$),
the case we have considered. 
In any case this effect can only partially compensate the suppression effect mentioned in the previous  point.
\item Another possible effect leading to a strong enhancement can come from density fluctuations if 
$\delta T/\bar{T} \gtrsim 1/(\beta/H_\star)$ \cite{Jinno:2021ury}.  From the reported results, the enhancement might
be up to an order of magnitude.  However, there are no specific calculations and at the moment such an effect should be 
regarded as potential. 
\end{itemize}
We can conclude that, within current knowledge, Eq.~(\ref{omegasw}) should be regarded as an upper bound of the
GW spectrum from first order phase transitions in the dark sector, likely for values $\alpha < 0.6$ from existing 
numerical simulations. Even for higher values of $\a$ there is currently no real reason to think there can be a strong enhancement,
rather a suppression, except for the hope that turbulence might become dominant and produce $\xi \gg 1$.  For this reason
in the following we will show results using Eq.~(\ref{omegasw}) as a plausible upper bound.  We will comment again in the 
final section about the possibility to evade such an upper bound.

\subsection{Results}

First of all we have produced scatter plots in the plane $\beta/H_\star$ versus $\alpha$ 
over the four parameters $v'_0$, $M'$, $\lambda'$, $C$ and for the three values $N'=1,3,5$.
The results are shown in the three panels in Fig.~1. The shadowed regions for $\alpha > 0.6$ 
indicate that in this regime there are no firm results from numerical simulations and for this reason 
we do not show benchmark points for such high values of $\alpha$. 
We also highlight benchmark points for which we show the GW spectrum in Fig.~2 and Fig.~3
for values of the parameters showed in the two tables.
\begin{figure}
	\centering
	\includegraphics[width=0.99\textwidth]{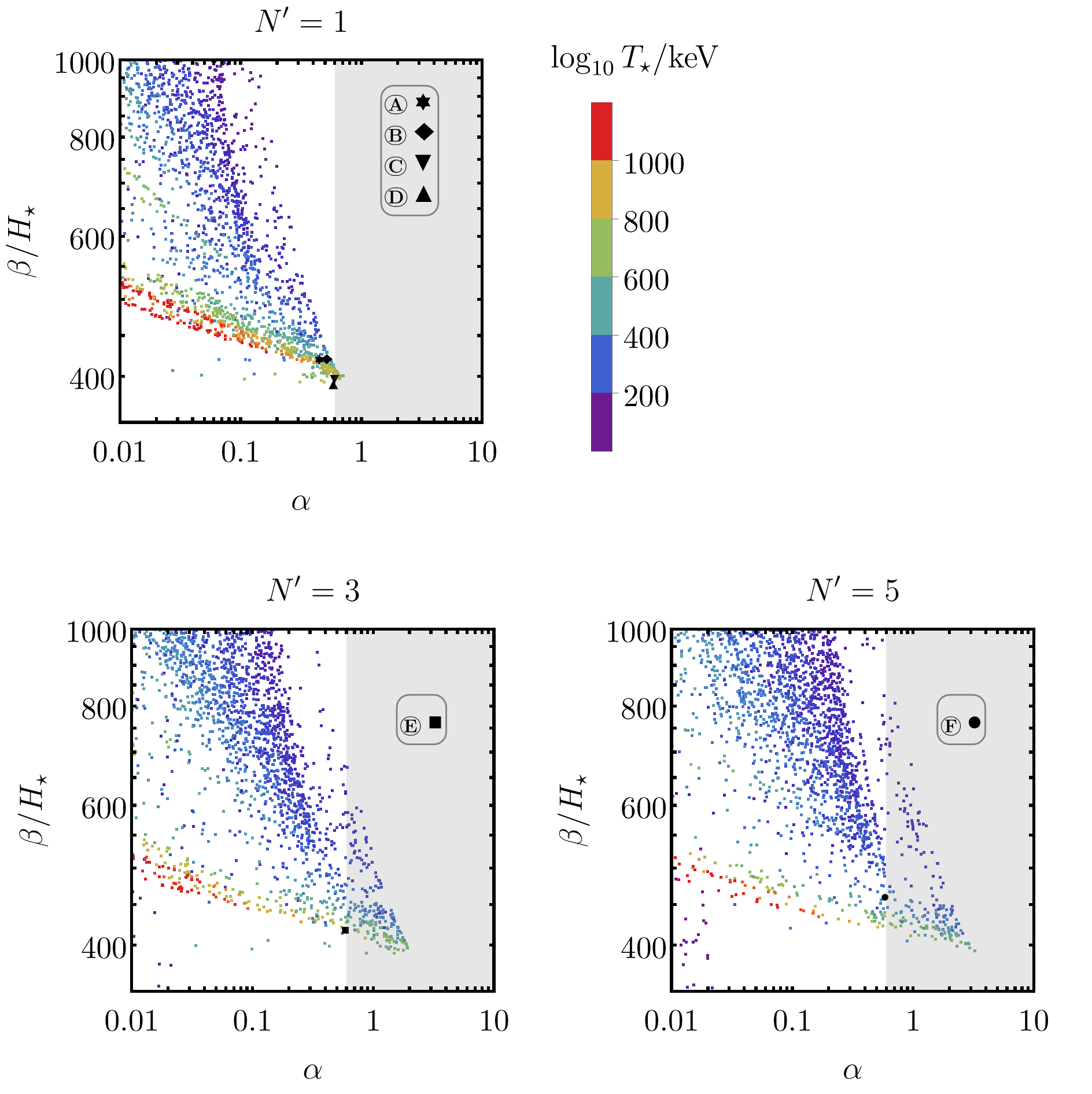} \\
	\caption{Scatter plot in the plane $\beta/H_\star$ versus $\alpha$ 
	over the four parameters $v'_0$, $M'$, $\lambda'$, $C$ and for the three values $N'=1,3,5$ corresponding
	to the three panels. The shadowed region indicates that for $\alpha \gtrsim 0.6$ we do not have
	a reliable expression for the GW spectrum. In the first panel, for $N' =1$, the diamond, lower and upper triangles
	indicate the three benchmark points in Table 1. The diamond in the first panel, the square in the second panel
	and the circle in the third panel indicate the three benchmark points in Table 2.}
	\label{fig:fig1}
\end{figure}

\begin{table}
\centering
{\renewcommand{\arraystretch}{1.2}
\begin{tabular}{c@{\hskip 0.1in}c@{\hskip 0.1in}c@{\hskip 0.1in}c@{\hskip 0.1in}c@{\hskip 0.1in}c@{\hskip 0.1in} c@{\hskip 0.1in}c@{\hskip 0.1in}c@{\hskip 0.1in}c@{\hskip 0.1in}c@{\hskip 0.1in}c@{\hskip 0.1in}c}
    \toprule
B.P. & $N'$  & $\lambda'$ & $v'_0$/keV & $M'$/keV & $C$/keV & $\alpha$ & $\alpha_{\nu{\rm D}}$ & $\kappa_{\nu {\rm D}}$ & $\beta/H_\star$ & $T_{\star}$/keV & $v_{\rm w}$ & $\Upsilon$  \\
 \midrule
 \Circled{A} & 1 & $0.0013$ & $54.85$  & $16.08$ & $0.96$ & $0.45$ & $2.06$ & $0.74$ & $423.93$ & $276.70$ & $0.96$ & $0.014$\\
 \Circled{B} & 1 & $0.001$ & $71.0$  & $20.0$ & $0.75$ & $0.52$ & $2.40$ & $0.74$ & $424.0$ & $240.58$ & $0.97$ & $0.013$\\
 \Circled{C} & 1 & $0.001$ & $83.0$  & $23.0$ & $1.70$ & $0.60$ & $2.62$ & $0.75$ & $399.73$ & $515.11$ & $0.97$ & $0.013$ \\
  \Circled{D} & 1 & $0.001$ & $144.0$  & $40.0$ & $3.0$ & $0.59$ & $2.56$ & $0.75$ & $393.63$ & $888.35$ & $0.97$ & $0.013$ \\
\bottomrule
\end{tabular}
}
\caption{Benchmark points for GW signals from first order phase transition of $\phi'$ for $N'=1$. }
\label{table:BP1}
\end{table}

\begin{figure}
	\centering
	\includegraphics[width=0.99\textwidth]{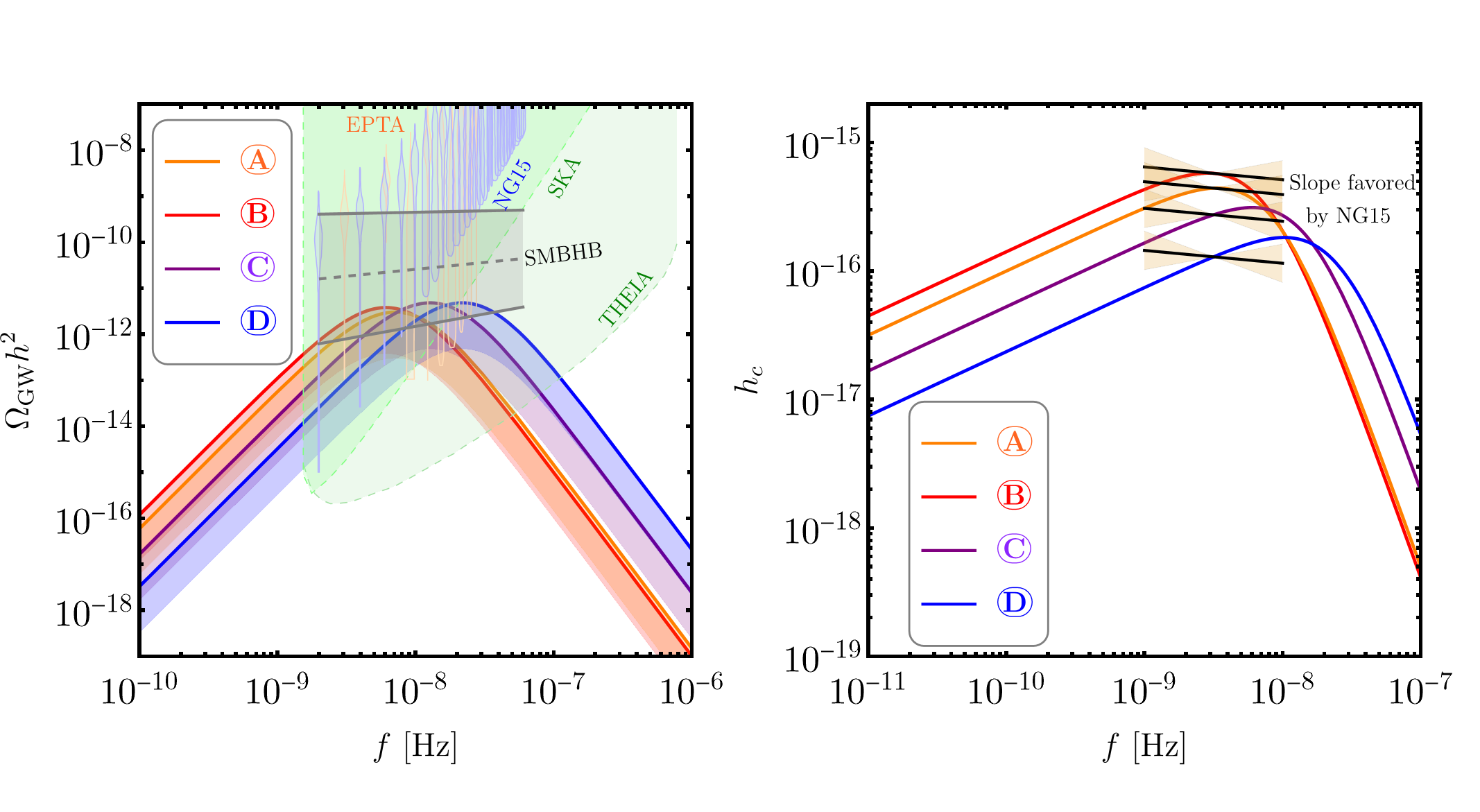}
	\caption{\emph{Left: }GW spectrum at NANOGrav for $N'=1$ and different $\alpha$. \emph{Right:} Strain spectrum compared to best fit from NANOGrav 15-yr data. Benchmark points are given in Table \ref{table:BP1}.}
	\label{fig:fig1}
\end{figure}

\begin{table}
\centering
{\renewcommand{\arraystretch}{1.2}
\begin{tabular}{c@{\hskip 0.1in}c@{\hskip 0.1in}c@{\hskip 0.1in}c@{\hskip 0.1in}c@{\hskip 0.1in}c@{\hskip 0.1in} c@{\hskip 0.1in}c@{\hskip 0.1in}c@{\hskip 0.1in}c@{\hskip 0.1in}c@{\hskip 0.1in}c@{\hskip 0.1in}c}
    \toprule

  B.P. & $N'$  & $\lambda'$ & $v'_0$/keV & $M'$/keV & $C$/keV & $\alpha$ & $\alpha_{\nu{\rm D}}$ & $\kappa_{\nu {\rm D}}$ & $\beta/H_\star$ & $T_{\star}$/keV & $v_{\rm w}$ & $\Upsilon$  \\
 \midrule
 \Circled{B} & 1 & $0.001$ & $71.0$  & $20.0$ & $0.75$ & $0.52$ & $2.40$ & $0.74$ & $424.0$ & $240.58$ & $0.97$ & $0.013$\\
 \Circled{E} & 3 & $0.001$ & $129.0$  & $29.0$ & $1.20$ & $0.59$ & $2.09$ & $0.71$ & $420.93$ & $262.61$ & $0.96$ & $0.013$\\
  \Circled{F} & 5 & $0.0013$ & $86.7$  & $14.18$ & $0.87$ & $0.59$ & $1.87$ & $0.69$ & $463.13$ & $218.65$ & $0.96$ & $0.012$\\
\bottomrule
\end{tabular}
}
\caption{Benchmark points for phase transition of $\phi'$ with  $N' = 1 ,3 ,5$, respectively. }
\label{table:BP2}
\end{table}

\begin{figure}
	\centering
	\includegraphics[width=0.99\textwidth]{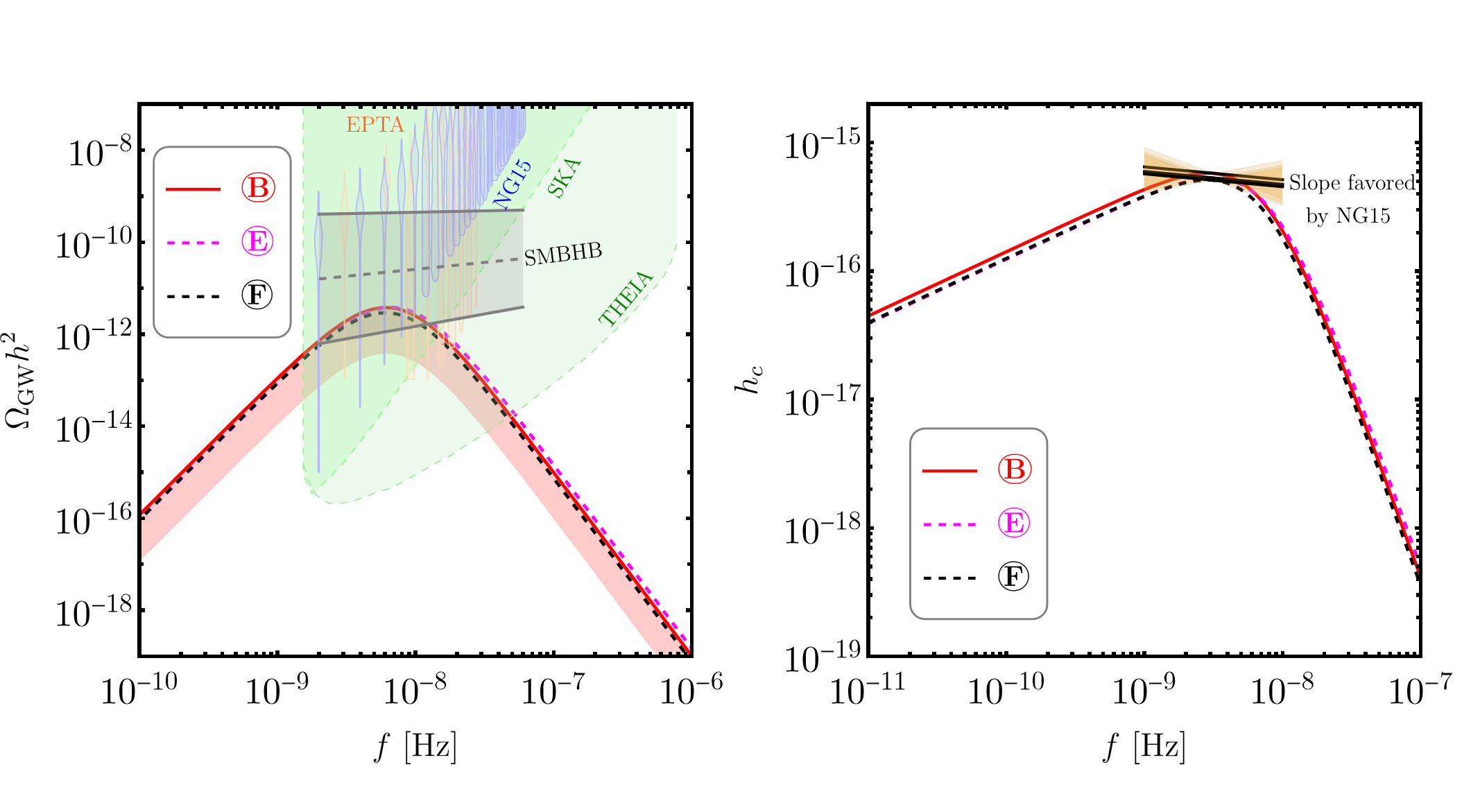}
	\caption{\emph{Left: }GW spectrum at NANOGrav with different $N'$. \emph{Right:} Strain spectrum compared to best fit from NANOGrav 15-yr data. Benchmark points are given in Table \ref{table:BP2}.}
	\label{fig:fig2}
\end{figure}

The left panel of Fig.~\ref{fig:fig1} shows four GW spectra, corresponding to the four benchmark points in Table 1, 
peaking in the frequency range probed by NANOGrav for $N'=1$. The peak amplitude of the signals are comparable, while the peak frequency shifts. 
In Fig.~\ref{fig:fig2}, we show two more benchmark points \Circled{E} and \Circled{F}, for $N'=3$ and $N'=5$ respectively. The resulting spectra is very similar to the benchmark point \Circled{B}, showing that the maximum  GW signal that can be achieved in this model in the NANOGrav frequencies is somewhat independent of $N'$. In these plots we have shown the sensitivity of some interferometers (SKA \cite{Janssen:2014dka}, THEIA \cite{Theia:2017xtk}) in the relevant frequency range with green shaded regions, and the recent NANOGrav \cite{NANOGrav:2023gor, NANOGrav:2023hvm} and EPTA \cite{EPTA:2023sfo, EPTA:2023fyk} results with blue and orange violins. 
These represent the symmetrical
representations of the 1D marginalized posterior probability density distributions of the GW energy density at each sampling frequency of the NANOGrav 15-yr and EPTA \cite{EPTA:2023fyk} data, respectively. 
We have also shown the baseline signal expected from supermassive black hole binaries (SMBHB), modeling their GW spectrum as a power-law fit following Ref.~\cite{NANOGrav:2023hvm}, where the dashed line shows the best fit and the bands correspond to $2\sigma$ deviations. Our model predicts larger amplitude than the worst case scenario of the baseline SMBHB model.

In the right panel of Figs.~\ref{fig:fig1} and \ref{fig:fig2} we show the dimensionless strain $h_c(f)$ of the GW signals, given by
\begin{align}
    h_c (f) = \sqrt{\frac{3 \mc H_0^2\ \Omega_{\rm GW}(f)}{2\pi^2 f^2}},
\end{align}
where $\mc H_0 \approx 68$ km/s/Mpc is today's Hubble rate. We compare the results with the spectral slope $\beta = {d\log h_c(f)}/{d \log f}$ modeling the NANOGrav strain spectrum with a simple power law of the form $h_c(f) = A_{\rm GW} (f/f_{\rm PTA})^\beta$. Expressing $\beta$ in terms of another parameter $\gamma_{\rm GW} = 3-2\beta$, the $1\sigma$ fit to  NANOGrav 15-yr data gives $\gamma_{\rm GW} \simeq 3.2 \pm 0.6$ around $f\sim 1/(10 \rm yr)$ \cite{NANOGrav:2023gor}. This favorable range is shown with bands superimposed on our strain plot. We find that the spectral tilt of the phase transition signal is in tension in some range of the frequency band probed by NANOGrav. 
\section{Final remarks}

Let us draw some final remarks on the results we obtained and how these 
can be further extended and improved. 

\begin{itemize}
\item Our results are compatible with those presented in \cite{DiBari:2021dri}. The differences
can be mainly understood in terms of the different expression we are using to describe the GW spectrum
from sound waves, the Eq.~(\ref{omegasw}). This supersedes the expression used in \cite{DiBari:2021dri}
based on \cite{Caprini:2015zlo}. The suppression factor taking into account the shorter duration of the 
stage of GW production compared to the duration of the phase transition is somehow compensated by the
fact that the new expression we are using is extended to higher values of $\alpha$. However, our description
of bubble velocity in terms of Jouguet solutions should be clearly replaced by a more advanced one taking into account friction
though we expect slight changes since the GW spectrum scales just linearly with $v_{\rm w}$. 
Another important difference is that  compared to \cite{Caprini:2015zlo} the peak frequency 
is more than halved for the same values of all relevant parameters such as $T_\star$.\footnote{This
is because the value of the coefficient $\kappa$ in \cite{Caprini:2015zlo} is taken $2/\sqrt{3} \simeq 1.2$.} 
This explains why we obtain higher values of $T'_\star \sim 100 \, {\rm keV}$ for the peak value to
be in the nHz range spanned by the NANOGrav signal. Also, notice that we have improved the calculation of the GW
spectrum taking properly into account the different temperature of the dark sector and calculating the efficiency
in terms of $\alpha_{\nu{\rm D}}$ rather than $\alpha$.
\item The peak amplitude we find is at most $h^2\Omega_{\rm gw}(f) \sim 10^{-11}$ at the NANOGrav frequencies
and cannot reproduce the whole NANOGrav signal. However, it can  help the contribution
from SMBHBs to improve the fit, one of the two options for the presence of new physics 
suggested by the NANOGrav collaboration analysis. This is certainly sufficient to make our results 
interesting, also considering that the model we studied is independently motivated by the cosmological tensions. 
Clearly, it would be interesting to perform a statistical analysis to find the best fit parameters in our model
and to quantify the statistical significance. 
\item The possibility to have a higher peak amplitude, corresponding to $\xi \gg 1$, 
cannot be excluded but from current results from numerical simulations it seems unlikely. We just notice that
increasing the value of $N'$, values of $\alpha$ higher than 0.6 are possible and since firm predictions are
missing for such strong first order transitions, one cannot  exclude large enhancement coming, for example, 
from not yet understood contribution from turbulence. A specific 
account of the effect of small fluctuations within our model might also offer potentially a way to obtain $\xi \gg 1$. 
\item The values $T'_{\star} \sim 100 \, {\rm keV}$ that we found in our solutions that 
enter the NANOGrav frequency range, imply $\Delta N_{\nu} \simeq 0.4$ 
at the time of nucleosynthesis, for $N' =1$. This is in marginal agreement with the constraint 
Eq.~(\ref{ubDNeetnuc}) from primordial deuterium measurements but it can be fully reconciled just simply
assuming extra degrees of freedom in the dark sector ($\D g \neq 0$). On the other hand, such an amount of dark radiation 
at the time of nucleosynthesis might be even beneficial to solve a potential deuterium problem. Actually if such 
a deviation from SBBN should be confirmed, this would provide quite a strong support to the model. 
Moreover, as discussed, the same amount of dark radiation
at recombination can ameliorate the cosmological tensions. In this respect a dedicated analysis within our model would 
be certainly desirable.  
\item One could think to explore a scenario with  $T'_{\star} \gg 1\,{\rm MeV}$, with a massive majoron $J'$  
quickly decaying before big bang nucleosynthesis thus avoiding all cosmological constraints \cite{Fairbairn:2019xog,Bringmann:2023opz}. This would require an extension
of the model introducing explicit symmetry breaking terms giving mass to $J'$. However, it has been noticed 
that the introduction of these terms leading to a majoron mass larger than about 1 eV would
actually jeopardise the occurrence of a first order phase transition \cite{DiBari:2021dri}. 
For this reason this possibility does not seem viable, since the decay rate of extremely ultra-relativistic 
particles is strongly suppressed.  
\item Finally, let us comment on the possibility to add to the tree level potential in Eq.~(\ref{Vzero2})
a usual Higgs portal interaction of the form $\lambda_{\Phi\phi'} |\Phi|^2 \, |\phi'|^2$. This is not
forbidden by the $U(1)_L$ symmetry and it is potentially interesting since one could directly 
consider the Higgs  as the auxiliary scalar field 
needed to enhance the strength of the $\phi'$ phase transition instead of $\phi$, making the model
more minimal. However, it is easy to see that such a possibility is excluded by the constraints
on the Higgs invisible decay width \cite{ATLAS:2018bnv,CMS:2018yfx} 
that place an upper bound $\lambda_{\Phi\phi'} \lesssim 0.03$ 
\cite{Bonilla:2015uwa,Addazi:2019dqt}. If we write the Higgs potential
as $V_0^{\rm SM}(\Phi) = -\mu^2_{\rm ew}\,|\Phi|^2 + \lambda_{\rm ew}|\Phi|^4$,
then $v_{\rm ew} = - \mu^2_{\rm ew}/(2\lambda_{\rm ew}) \simeq 174\,{\rm GeV}$.
Expanding $\Phi$ about the electroweak VEV, one has $|\Phi| = (0, v_{\rm ew} + h/\sqrt{2})^T$, where
$h$ is the Higgs boson field with mass $m_{h} = 2\sqrt{\lambda_{\rm ew}}\,v_{\rm eq} \simeq 125\,{\rm GeV}$,
so that one has $\lambda_{\rm ew} = m^2_h/(4\, v^2_{\rm ew}) \simeq 0.13$. After electroweak symmetry breaking, 
the Higgs portal term would give an additional contribution to the zero temperature cubic term 
$C$  given by  
\be
C_{\Phi} = {\lambda_{\Phi\phi'}^2\,v_0'\over 2\lambda_{\rm ew}} \lesssim 0.35\,{\rm keV}\, {v_0'\over 100\,{\rm keV}} \,  .
\ee
If this is compared to the values of $C$ obtained for the benchmark points, it seems that this contribution
is sub-dominant. However, since it is only marginally sub-dominant, it would be certainly interesting to explore in more
detail the very attractive possibility that the NANOGrav signal might have some connection with a potential
contribution of majorons to the Higgs invisible decay width that should be discovered at colliders. 
\end{itemize}
In conclusion the split majoron model is an appealing possibility to address part of the NANOGrav signal and the cosmological
tensions, including, potentially, the deuterium problem. Moreover, it can have connections
with other different phenomenologies. In any case it is certainly a clear example of how, with the evidence of a GW cosmological background  from NANAOGrav,  GWs have opened a new era in our quest of new physics.
This should certainly alleviate the regret for the non-evidence of new physics at the LHC (at least so far).

\vspace{-1mm}
\subsection*{Acknowledgments}

We acknowledge financial support from the STFC Consolidated Grant ST/T000775/1.
We also would like to thank David Weir for many useful discussions.

\end{document}